# Formation of k-space indirect magnetoexcitons in double-quantum-well direct gap heterostructures

A. A. Gorbatsevich         I. V. Tokatly

Moscow Institute of Electronic Engineering, Zelenograd, 103498, Russia

**Abstract**

Spectrum of excitons in double-quantum-well structure is calculated in a tilted magnetic field. It is shown that spectrum becomes asymmetric in quasimomentum if a transverse in respect to growth direction component of magnetic field is nonzero. A transition from **k**-space direct exciton ground state to **k**-space indirect one accompanied by sharp quenching of photoluminescence is described.

# 1   Introduction

Study of magnetoexcitons in Multiple-Quantum-Well-Structures (MQWS) is a topic of permanent interest. It provides information about the interplay of interelectron interactions and size- and magnetic- quantization phenomena. Majority of these studies deals with magnetic field parallel to growth direction $\mathbf{B}_{par}$. Energy spectrum of Quantum Well (QW) states in this case is fully quantized and the properties of exciton states are in many respects similar to those of 2D systems. However an additional degrees of freedom connected with the interwell Coulomb interaction and with the possibility of interwell carrier transfer make the situation more complex. In coupled QWs both direct excitons with electron and hole in the pair located in the same well and indirect excitons comprised of electron and hole from adjacent wells are possible. In asymmetric MQWS (in coupled QWs of second type) indirect excitonic state can become a ground state. Recently the crossover from direct to indirect excitonic states was observed under the variation of parallel magnetic field $\mathbf{B}_{par}$ [1]. This crossover is accompanied by rapid variation of luminescence intensity which is substantially reduced for indirect exciton due to smaller electron and hole wave-functions overlap. The origin of this crossover is the variation of direct excitonic binding energy $E_d$ which continuously increases in high magnetic field as $E_d \propto e^2/\lambda$ ($\lambda$ - is magnetic length) and saturates for indirect exciton at value $E_{id} \propto e^2/d$ ($d$ - is interwell separation).

In the present paper we investigate much less studied situation with magnetic field ($\mathbf{B}_{perp}$) perpendicular to growth direction. Here as we show below variation of the magnetic field can also produce rapid variation of luminescence intensity. However contrary to the case of $\mathbf{B}_{par}$ luminescence intensity is reduced at high fields. The reason is that $\mathbf{B}_{perp}$ makes an indirect in r-space exciton also indirect in $\mathbf{k}$-space which results in further suppression of radiative recombination.

Properties of MQWS exciton spectrum in $\mathbf{B}_{perp}$ are generically coupled to the properties of one-particle (electron and hole) energy spectrum. The latter was studied in a number of theoretical [2, 3, 4] and experimental [5, 7] publications. Peculiarities of one-particle spectrum transformation in magnetic field are connected with the relation between center-of-Landau-orbit coordinate $z_0$ and in-plane momentum $k_y = z_0/\lambda^2$. Carriers located in different QWs separated by a distance d acquire a relative shift in momentum space $\delta k = d/\lambda^2$. Manifestation of this shift were observed in experiments on interwell resonant tunneling where it influences the momentum conservation condition and essentially modifies tunneling I-V curve [5]. Spectrum modifications in the in-plane magnetic field was probed by the in-plane magnetoresistance measurements [6]. In AMQWS centers of electron and hole envelopes don't coincides even in the same QW. In $\mathbf{B}_{perp}$ real space shift of electron and hole envelopes generates a relative shift in momentum space of electron and hole bands. Hence a direct gap semiconductor can become an indirect one which was experimentally observed in [7]. If the interwell coupling is strong and the coherence is established over the whole MQWS the system can be characterized by energy spectrum asymmetric in momentum [4]. Such asymmetry can result in observable macroscopic phenomena: anomalous photogalvanic [4] and magnetoelctric [8] effects. The former one was observed in [9].

Variation of exciton binding energy in magnetic field $\mathbf{B}_{par}$ was calculated in [10, 11]. Systematic study of exciton energy spectrum became possible after the remarkable observation of Gor'kov and Dzyaloshinski [12] who introduced an analog of conserving momentum for center of mass motion. Based on these results spectrum of 2D excitons in $\mathbf{B}_{par}$ was calculated [13]. In the present paper we calculate using a tight binding approach exciton spectrum in a tilted magnetic field directed at an arbitrary angle with respect to QW growth direction. The structure of the paper is the following. In Sec.2 we introduce the one-particle tight-binding basis, discuss an appropriate choice of vector-potential gauge, explicitly construct one-particle asymmetric spectrum and estimate the renormalization of tunneling matrix elements in magnetic field. In Sec.3 the tight-binding approach to the exciton problem in MQWS is formulated which makes it possible to decouple relative and center-of-mass motions at arbitrary orientation of magnetic field. In Sec.4 exciton energy spectrum is calculated which in AMQWS is asymmetric in momentum. The degree of asymmetry can be controlled by external electric field parallel to growth direction and/or $\mathbf{B}_{perp}$. It is shown that under sufficiently strong real space asymmetry exciton state with nonzero total momentum becomes a ground state. In optical experiment it will result in photoluminescence quenching.

## 2  Single-particle spectrum of DQWS in the presence of in-plane magnetic field

The effective mass hamiltonian describing electron (hole) in heterostructure in the presence of crossed magnetic field $\mathbf{B}$ and electric field $\mathbf{F}$ is determined by the following expression

$$H_{e(h)} = \frac{1}{2m_{e(h)}} \left( \hat{\mathbf{k}}_{e(h)} \pm \frac{e}{c} \mathbf{A}(\mathbf{r}_{e(h)}) \right)^2 + U_{e(h)} \left( z_{e(h)} \right) \mp e\mathbf{r}_{e(h)} \mathbf{F} \qquad (1)$$

where $\hat{\mathbf{k}}_{e(h)}$ - is momentum operator, $m_{e(h)}$ is electronic (hole) mass, $A(\mathbf{r})$ is vector potential and $U_{e(h)}(z)$ is heterostructure potential ($z$-axis is in growth direction). The upper and lower signs in (1) correspond to electron and hole hamiltonians. Let only the in-plane component of magnetic field to be nonzero and electric field $\mathbf{F}$ to be applied along growth direction. Namely let us $\mathbf{B} = \mathbf{e}_x B_x$ for definiteness and $\mathbf{F} = \mathbf{e}_z F$ ($\mathbf{e}_x$ and $\mathbf{e}_z$ are the unit vectors along $x$ and $z$ axis correspondingly). We consider for simplicity symmetrical DQWS. The asymmetry is modelled by external electric field $\mathbf{F}$. Such asymmetry corresponds to coupled QWs of the second type. We choose Landau gauge for vector potential $\mathbf{A}$

$$\mathbf{A}(\mathbf{r}) = \mathbf{e}_z B_x y.$$

the next step is introduction of magnetic field dependent tight-binding basis functions $|e(h), j\rangle$ ($j = 1, 2$ is QW number)

$$|e(h), j\rangle = \varphi_{e(h)}^{j}\left(z_{e(h)}\right) \exp\left\{\mp i \frac{z_{e(h)} - z_j}{\lambda_x^2} y_{e(h)}\right\}. \quad (2)$$

Where $\varphi_{e(h)}^{j}\left(z_{e(h)}\right)$ is the tight-binding basis function describing electron (hole) in the $j$-th well in the absence of external fields. Such choice of basis functions assumes that size-quantization energy is the largest energy scale and the relation

$$\Delta E_n \gg \hbar \omega_0$$

where $\Delta E_n$ - is energy separation between size-quantized levels in each well and $\omega_0 = eB/mc$ - is the cyclotronic frequency. The phase factor in (2) describes the in-plane magnetic field effects ($\lambda_x^{-2} = eB_x/c$ is magnetic length) and $z_j$ is the coordinate of the $j$-th QW center. In the absence of external fields ($B_x = F = 0$) the two-level tight-binding hamiltonian takes the form of $2 \times 2$ matrix

$$H = \varepsilon_{e(h)}^{(0)} + \frac{\hat{\mathbf{k}}_\perp^2}{2m_{e(h)}} - t_{e(h)}^{(0)} \sigma_x \quad (3)$$

here $\varepsilon_{e(h)}^{(0)}$ and $t_{e(h)}^{(0)}$ are intrawell energy levels and interwell tunneling matrix element correspondingly, $\hat{\mathbf{k}}_\perp$ is in-plane momentum operator and $\sigma_x$ is Pauli matrix. According to (1) and (2) the tunneling probability acquires the phase multiplicator in the case of nonzero magnetic field. The absolute values of $\varepsilon_{e(h)}$ and $t_{e(h)}$ are renormalized by external fields either. As a result the hamiltonian takes the form

$$H = \begin{bmatrix} \varepsilon_{e(h)} + \frac{\hat{\mathbf{k}}_\perp^2}{2m_{e(h)}} \pm edF & -t_{e(h)} \exp\left\{\mp i \frac{dy}{\lambda_x^2}\right\} \\ -t_{e(h)} \exp\left\{\pm i \frac{dy}{\lambda_x^2}\right\} & \varepsilon_{e(h)} + \frac{\hat{\mathbf{k}}_\perp^2}{2m_{e(h)}} \mp edF \end{bmatrix} \quad (4)$$

where $d$ is interwell distance and $\varepsilon_{e(h)}$ and $t_{e(h)}$ correspond to the renormalized values of intrawell energy and interwell tunneling matrix element. Using perturbation theory one can obtain

$$\varepsilon_{e(h)} = \varepsilon_{e(h)}^{(0)} + \frac{\omega_x}{2} \frac{\langle \varphi_{e(h)}^{1(2)} | z^2 | \varphi_{e(h)}^{1(2)} \rangle}{\lambda_x^2} \quad (5)$$

$$t_{e(h)} = t_{e(h)}^{(0)} - \frac{\omega_x}{2} \frac{\langle \varphi_{e(h)}^{1} | z^2 | \varphi_{e(h)}^{2} \rangle}{\lambda_x^2} \quad (6)$$

where $\omega_x = 1/m_{e(h)} \lambda_x^2$ is the cyclotronic frequency corresponding to $B - x$-component of the magnetic field. The increasing of $\varepsilon_{e(h)}$ corresponds to the usual diamagnetic shift $\delta \varepsilon(B) \propto \omega_x \left(\frac{L_w}{\lambda_x}\right)^2$ ($L_w$ is the well thickness). The decreasing of tunneling probability (6) is due to the squeezing of wave functions by in-plane magnetic field. The simple estimate of the second term in (6) gives $\frac{\delta t}{t} \propto -\left(\frac{L_w d}{\lambda_x^2}\right)^2$. We neglect in (4), (5) the intrawell Stark effect. Taking into account of this effect leads to the addition of quadratic Stark shift $\delta\varepsilon(F) \propto (eFL_w)^2/\varepsilon^{(0)}$ in (5) and does not change qualitative results.

The space inhomogeneous phase factor in (4) can be eliminated by appropriate gauge transformation

$$H = UHU^+ \quad ; \quad U = \exp\left\{\pm i \frac{dy}{\lambda_x^2}\sigma_z\right\}$$

After the transformation we get

$$H = \varepsilon_{e(h)} \pm eFd\sigma_z + \frac{1}{2m_{e(h)}}\left(\hat{\mathbf{k}}_\perp \pm \mathbf{e_y}\frac{d}{2\lambda_x^2}\sigma_z\right)^2 - t_{e(h)}\sigma_x \tag{7}$$

Energy spectrum corresponding to (7) consists of two parabolas shifted in momentum space

$$E_{e(h)}(\hat{\mathbf{k}}_\perp) = \varepsilon_{e(h)} + \frac{1}{2m_{e(h)}}\left(\hat{\mathbf{k}}_\perp^2 + \left(\frac{d}{2\lambda_x^2}\right)^2\right) \pm \sqrt{\left(\frac{k_y d}{2m_{e(h)}\lambda_x^2} + eFd\right)^2 + t_{e(h)}^2} \tag{8}$$

The ground branch of the electron and hole dispersion law (8) has two minima approximately located at the points $k_y \approx \pm d/2\lambda_x^2$. In the presence of electric field (or more generally in the presence of any kind of QW asymmetry) single particle spectrum becomes asymmetric in momentum. So the indirect energy gap is produced by crossed external fields [7] in the initially direct gap semiconductor. The analytical expression for energy spectrum (8) was derived under the special choice of basis functions (2). However the appearance of spectrum asymmetry in momentum in transverse magnetic field is a very general result connected with the violation of both time and space parity in AMQWS in magnetic field. In such structures a linear in momentum $\mathbf{k}$ term in energy is allowed by the symmetry:

$$\Delta E \propto [\mathbf{H} \times \mathbf{l}]\mathbf{k}$$

here $\mathbf{l}$ - is a polar vector lying in growth direction and describing structure asymmetry.

There are a number of possible excitonic states in the electron-hole system with multiextrema spectrum (8). In the next section we develop the regular procedure based on the generalization of the tight-binding approach on the two-body (electron-hole) problem for describing all of exciton states in the whole range of momenta.

# 3 Hamiltonian of the system and basic equations

We consider DQW structure in electric field $\mathbf{F}$ parallel to growth direction $z$ and magnetic field $\mathbf{B}$. Let $\varphi$ to be the angle between magnetic field direction and $z$-axis , and for definiteness we assume that vector $\mathbf{B}$ lies in $z - x$ plane and has two components $B_x = B \sin \varphi$ and $B_z = B \cos \varphi$. Initial electron-hole two-particle hamiltonian in effective-mass approximation has usual form:

$$H = \frac{1}{2m_e}\left(\hat{\mathbf{k}}_e + \frac{e}{c}\mathbf{A}_e\right)^2 + \frac{1}{2m_h}\left(\hat{\mathbf{k}}_h - \frac{e}{c}\mathbf{A}_h\right)^2 + U_e(z_e) + U_h(z_h) + V_c(\mathbf{r}_e - \mathbf{r}_h) \quad (9)$$

where $V_e(\mathbf{r}_e - \mathbf{r}_h)$ is the Coulomb potential. For convinience we use in this section Landau gauge for $x$-component of magnetic field and circular gauge - for $z$-component. In this gauge the vector-potential $\mathbf{A}$ has the form:

$$\mathbf{A}(\mathbf{r}) = \mathbf{e}_z y B_x + \frac{1}{2}\mathbf{e}_z \times \mathbf{r} B_z \quad (10)$$

where $\mathbf{e}_z$ is the unit vector lying in growth direction $z$.

To describe excitonic effects we introduce the "two-particle" tight-binding basis with the following four basic functions $|i,j\rangle$ ($i,j$ are QW numbers):

$$|ij\rangle = |e,i\rangle|h,j\rangle \quad (11)$$

The function $|e(h), j\rangle$ is determined by the expression (2) and describes the electron (hole) located in $j$-th well in the absence of tunneling. Phase factor in (4) allows for the transverse component of the magnetic field $B_x$ and $\lambda_x^{-2} = eB_x/c$ is the magnetic length corresponding to $B_x$. The introduction of the basis (11) makes it possible to separate parallel (in respect to $z$-direction) and transverse motion and to derive matrix Schroedinger equation for wave functions depending only on the transverse coordinates $\mathbf{r}_{e,h}$.

Let the left QW (number 1) be narrow and the right one (number 2) be wide so that the single-particle energy levels in the wells are:

$$\varepsilon_{e,h}^{1,2} = \overline{\varepsilon}_{e,h} \pm \Delta_{e,h}$$

Denote the electron and hole interwell tunneling matrix elements as $t_e$ and $t_h$. Projecting the hamiltonian on the basis (11) we obtain the following system of equations

$$\left[\hat{h} - V_{11}(\mathbf{r}_e - \mathbf{r}_h) + \Delta_1\right]\chi_1 - t_h \exp\left(-i\frac{dy_h}{\lambda_x^2}\right)\chi_3 - t_e \exp\left(i\frac{dy_e}{\lambda_x^2}\right)\chi_4 = \varepsilon\chi_1$$

$$\left[\hat{h} - V_{22}(\mathbf{r}_e - \mathbf{r}_h) - \Delta_1\right]\chi_2 - t_e \exp\left(-i\frac{dy_e}{\lambda_x^2}\right)\chi_3 - t_h \exp\left(i\frac{dy_h}{\lambda_x^2}\right)\chi_4 = \varepsilon\chi_2 \quad (12)$$

$$\left[\hat{h} - V_{12}\left(\mathbf{r}_e - \mathbf{r}_h\right) + \Delta_1\right]\chi_3 - t_h \exp\left(i\frac{dy_h}{\lambda_x^2}\right)\chi_1 - t_e \exp\left(-i\frac{dy_e}{\lambda_x^2}\right)\chi_2 = \varepsilon\chi_3$$

$$\left[\hat{h} - V_{21}\left(\mathbf{r}_e - \mathbf{r}_h\right) - \Delta_2\right]\chi_4 - t_e \exp\left(-i\frac{dy_e}{\lambda_x^2}\right)\chi_1 - t_h \exp\left(i\frac{dy_h}{\lambda_x^2}\right)\chi_2 = \varepsilon\chi_4$$

where $\chi_i$ are the components of the system wave-function in the basis (11). In (12) $\Delta_{1,2} = \Delta_e \pm \Delta_h$, $\overline{\varepsilon}_g = \overline{\varepsilon}_e + \overline{\varepsilon}_h$ and

$$\hat{h} = \frac{\hat{\mathbf{P}}_e^2}{2m_e} + \frac{\hat{\mathbf{P}}_h^2}{2m_h} + \overline{\varepsilon}_g \quad ; \quad \hat{\mathbf{P}}_{e,h} = \hat{\mathbf{k}}_{\perp,e,h} \pm \frac{e}{2c}\mathbf{e}_z \times \mathbf{r}_{e,h}B_z \tag{13}$$

The matrix elements of the Coulomb potential $V_{ij}(\mathbf{r}_e - \mathbf{r}_h)$ corresponding to interaction between an electron in $i$-th well and a hole in $j$-th well are

$$V_{ij} = \int dz_e dz_h \varphi_{i,e}^2(z_e) V_c(\mathbf{r}_e - \mathbf{r}_h) \varphi_{j,h}^2(z_h) \tag{14}$$

The next step is excitonic center of mass motion separation. We use the procedure similar to the one proposed in [12] and generalizes it to the case of multicomponent wave-function. Namely we search the solution of the system of equations (12) in the following form

$$(\chi_1, \chi_2, \chi_3, \chi_4) = (u_1(\mathbf{r}), u_2(\mathbf{r}), u_3(\mathbf{r})e^{-i\frac{dY}{\lambda_x^2}}, u_4(\mathbf{r})e^{i\frac{dY}{\lambda_x^2}}) \exp\left\{i\left(\mathbf{Q} + \frac{e}{2c}\mathbf{e}_z \times \mathbf{r}B_z\right)\mathbf{R}\right\} \tag{15}$$

where center-of-mass coordinate $\mathbf{R} = (\mathbf{r}_e m_e + \mathbf{r}_h m_h)/M$ ($M = m_e + m_h$) and coordinate of relative motion $\mathbf{r} = \mathbf{r}_e - \mathbf{r}_h$ are introduced. After substitution of wave function (15) in (12) we obtain matrix Schroedinger equation with the Hamiltonian

$$H = \begin{bmatrix} \hat{h}_{11}(\hat{\mathbf{q}}, \mathbf{Q}, \mathbf{r}) & 0 & -t_h(y) & -t_e(y) \\ 0 & \hat{h}_{22}(\hat{\mathbf{q}}, \mathbf{Q}, \mathbf{r}) & -t_e^*(y) & -t_h^*(y) \\ -t_h^*(y) & -t_e(y) & \hat{h}_{12}\left(\hat{\mathbf{q}}, \mathbf{Q} - \mathbf{e}_y\frac{d}{\lambda_x^2}, \mathbf{r}\right) & 0 \\ -t_e^*(y) & -t_h(y) & 0 & \hat{h}_{21}\left(\hat{\mathbf{q}}, \mathbf{Q} + \mathbf{e}_y\frac{d}{\lambda_x^2}, \mathbf{r}\right) \end{bmatrix} \tag{16}$$

$$\hat{h}_{11,22}(\hat{\mathbf{q}}, \mathbf{Q}, \mathbf{r}) = \hat{h}(\hat{\mathbf{q}}, \mathbf{Q}) - V_{11,22}(\mathbf{r}) \pm \Delta_1$$
$$\hat{h}_{12,21}(\hat{\mathbf{q}}, \mathbf{Q}, \mathbf{r}) = \hat{h}(\hat{\mathbf{q}}, \mathbf{Q}) - V_{12,21}(\mathbf{r}) \pm \Delta_2$$

Here $\hat{\mathbf{q}} = -i\partial/\partial \mathbf{r}$ is relative motion momentum operator, $\mathbf{Q}$ is center-of-mass momentum. Kinetic energy operator $\hat{h}(\hat{\mathbf{q}}, \mathbf{Q})$ takes the standard form in center-of-mass coordinate system:

$$\hat{h}(\hat{\mathbf{q}}, \mathbf{Q}) = \frac{\hat{q}^2}{2m} - \gamma\frac{\mathbf{e}_z \times \mathbf{r}}{2m\lambda_z^2}\hat{\mathbf{q}} + \frac{(\mathbf{e}_z \times \mathbf{r})^2}{8m\lambda_z^4} + \frac{\mathbf{e}_z \times \mathbf{r}}{M\lambda_z^2}\mathbf{Q} + \frac{\mathbf{Q}^2}{2M} \tag{17}$$

here $m = m_e m_h / M$ is the reduced mass and $\gamma = (m_e - m_h)/M$. If the transverse component of the magnetic field $B_x$ is nonzero then tunneling matrix elements $t_{e,h}$ acquire phase factors and become the functions of the coordinate $y$:

$$t_{e,h}(y) = t_{e,h} \exp\left(i \frac{dy}{\lambda_x^2} \frac{m_{h,e}}{M}\right)$$

The hamiltonian (16) acts on the 4-component wave-function $\Phi(\mathbf{r})$ determined in (15).

An important conclusion can be at once deduced from the general form of the hamiltonian (16) written the local "two-particle" basis (11) under $B_x \neq 0$. Center-of-mass momentum $\mathbf{Q}$ in the diagonal components of the hamiltonian corresponding to indirect in $r$-space excitons is shifted by $\delta \mathbf{Q} = \mathbf{e}_y \frac{d}{\lambda_x^2}$ (here d -is separation of electron and hole along $z$-axis). Hence under application of the transverse in respect to growth direction magnetic field indirect in real space exciton become also indirect in $\mathbf{Q}$-space. This momentum shift obviously results from the correspondence between charged particle center of orbit in the magnetic field $\mathbf{B} = \mathbf{e}_x B_x$ and $y$-component of the momentum $z_0 = \lambda_x^2 k_y$. Hence electron and hole separated in real space by the distance d are separated in momentum space by vector $\mathbf{e}_y d / \lambda_x^2$ as well.

It is straightforward to rewrite the hamiltonian (16) in the basis of its diagonal part eigen-functions. We assume that the solutions are known of the equations:

$$\left[\hat{h}(\hat{\mathbf{q}}, \mathbf{Q}) - V_{ij}(\mathbf{r})\right] \psi_{ij}(\mathbf{r}) = E_{ij}(\mathbf{Q}) \psi_{ij}(\mathbf{r}) \tag{18}$$

where $\hat{h}(\hat{\mathbf{q}}, \mathbf{Q})$ is determined in (20) and $V_{i,j}(\mathbf{r})$ - in (14). Using the substitution $\psi_{i,j} = e^{\frac{i}{2}\gamma \mathbf{r} \mathbf{Q}} \phi_{i,j}(\mathbf{r} - \mathbf{r_0})$ we obtain from (18) the following equations:

$$\left\{\frac{\hat{q}^2}{2m} - \gamma \frac{\mathbf{e}_z \times \mathbf{r}}{2m\lambda_z^2} \hat{\mathbf{q}} + \frac{\mathbf{r}^2}{8m\lambda_z^4} - V_{ij}(\mathbf{r} + \mathbf{r_0})\right\} \phi_{ij}(\mathbf{r}) = E_{ij}(\mathbf{Q}) \phi_{ij}(\mathbf{r}) \tag{19}$$

$$\mathbf{r_0}(\mathbf{Q}) = \lambda_z^2 \mathbf{e}_z \times \mathbf{Q}$$

It follows from the structure of equations (19) that eigen-energies $E_{i,j}(\mathbf{Q})$ - are even functions of the momentum $\mathbf{Q}$. Consider the case when tunneling mixes only ground excitonic states. Under this assumption reasonable approach is to consider 4-component basis comprising two real-space-direct and two real-space-indirect exciton wave functions. Moreover for simplicity we consider symmetrical DQW structure. The asymmetry is modeled by the external electric field $\mathbf{F}$ directed along the $z$ axis $\mathbf{F} = \mathbf{e}_z F$. In symmetrical DQW structure eugene functions $\psi_{ij}(\mathbf{r})$ and eugene energies $E_{i,j}(\mathbf{Q})$ are symmetrical in well indices. Assume the following notations:

$$\psi_{11} = \psi_{22} = \psi_d \quad ; \quad E_{11} = E_{22} = E_d$$

for wave function and energy of direct excitons and analogously

$$\psi_{12} = \psi_{21} = \psi_{id} \quad ; \quad E_{12} = E_{21} = E_{id}$$

- for indirect in **r**-space excitons.

In the chosen basis of 4 local excitonic states we obtain the system of 4 algebraic equations

$$H\Psi = (\varepsilon - \varepsilon_g)\Psi$$

with the hamiltonian

$$H = \begin{bmatrix} E_d(\mathbf{Q}) & 0 & -T_h & -T_e \\ 0 & E_d(\mathbf{Q}) & -T_e^* & -T_h^* \\ -T_h^* & -T_e & E_{id}\left(\mathbf{Q} - \mathbf{e}_y \frac{d}{\lambda_x^2}\right) + edF & 0 \\ -T_e^* & -T_h & 0 & E_{id}\left(\mathbf{Q} + \mathbf{e}_y \frac{d}{\lambda_x^2}\right) - edF \end{bmatrix} \quad (20)$$

The matrix elements $T_e$ and $T_h$ describing mixing of excitonic states due to electron and hole tunneling have the form:

$$T_{e,h} = t_{e,h} \int \exp\left(i \frac{dy}{\lambda_x^2} \frac{m_{h,e}}{M}\right) \psi_d^*(\mathbf{r})\psi_{id}(\mathbf{r}) d^2\mathbf{r} \quad (21)$$

## 4  Excitonic spectrum in external fields

Exciton energy momentum dependence is determined by the solution of dispersion equation:

$$det(\varepsilon - H) = 0 \quad (22)$$

where $H$ - $4 \times 4$ matrix determined in (20). Excitonic spectrum consists of 4 branches. These branches are generically coupled with two degenerate branches of direct excitons

$$\varepsilon_{1,2}(\mathbf{Q}) = E_d(\mathbf{Q}) \quad (23)$$

and two branches of indirect excitons

$$\varepsilon_{3,4}(\mathbf{Q}) = E_{id}\left(\mathbf{Q} \pm \mathbf{e}_y \frac{d}{\lambda_x^2}\right) \pm edF \quad (24)$$

The extrema of two latter branches are symmetrically shifted in **k**-space at $B_x = B\sin\varphi \neq 0$. Tunneling lifts the degeneracy at the intersection points of (23) and (24). In these points mixing of different excitonic states is most strong. Note that in the presence of electric field **F** excitonic spectrum becomes asymmetric in respect to $k_y$ direction.

Before starting to analyze exciton energy spectrum in details let us consider the influence of magnetic field on the elements of hamiltonian matrix (20).

The diagonal elements correspond to the ground state energy of direct and indirect excitons in the absence of tunneling. These energies are obtained from the solution of equation (19) and hence depend only upon the $z$-component of magnetic field $B_z$. From (19) it follows that $E(\mathbf{Q})$ is even function of the momentum $\mathbf{Q}$. Hence at small $\mathbf{Q}$ one can take

$$E_{d,id}(\mathbf{Q}) \approx E_{d,id}(0) + \frac{Q^2}{2M^*_{d,id}}$$

In particular at $B_z = 0$ it is evident that

$$E_{d,id}(0) = \varepsilon_{d,id} \quad ; \quad M^*_{d,id} = M \equiv m_e + m_h$$

In the latter formula $\varepsilon_{d,id}$ - is exciton binding energy in the absence of external fields. The inverse limit of strong field along $z$ axis is realized under the condition

$$a_{ex}/\lambda_z \gg 1 \qquad (25)$$

Here $a_{ex}$ is the minimum of direct and indirect exciton radii ($a_{ex} = min\{a_d, a_{id}\}$). In this case exciton dispersion law can be calculated treating Coulomb interaction $V_{ij}(\mathbf{r})$ as perturbation [13]

$$E_{ij}(\mathbf{Q}) \approx \frac{1}{2}\omega_c - \int d^2 r \phi_0^2(r) V_{ij}\left(\mathbf{r} + \lambda_z^2 \mathbf{e}_z \times \mathbf{Q}\right)) \qquad (26)$$

here $\omega_c = eB/Mc$, and $\phi_0(r)$ - is oscillator wave function

$$\phi_0(r) = \frac{1}{\sqrt{2\pi}\lambda_z} \exp\left(-\frac{r^2}{4\lambda_z^2}\right) \qquad (27)$$

For direct exciton binding energy we obtain

$$\frac{1}{2}\omega_c - E_d = \frac{e^2}{\lambda}\sqrt{\frac{\pi}{2}} e^{-\frac{1}{4}\lambda_z^2 Q^2} I_0\left(\frac{\lambda_z^2 Q^2}{4}\right) \approx \frac{e^2}{\lambda}\sqrt{\frac{\pi}{2}}\left(1 - \frac{1}{4}\lambda_z^2 Q^2\right) \qquad (28)$$

In (28) $I_\nu$ is Bessel function. For indirect exciton a simple analytic formula can be deduced in the limit of weakly coupled QWs ($d/\lambda_z \gg 1$)

$$E_{id} \approx \frac{1}{2}\omega_c - \frac{e^2}{\sqrt{d^2 + \lambda_z^4 Q^2}} \qquad (29)$$

In reverse limit $d/\lambda_z \ll 1$ (but under the condition (25)) we have $E_d(\mathbf{Q}) \approx E_{id}(\mathbf{Q})$.

So under the increasing magnetic field $B_z$ dispersion curves of direct and indirect excitons differs substantially. Indirect exciton appears to be much heavier than the direct one. In the limit $d/\lambda_z \gg 1$ e.g. it follows from (28), (29) that $M^*_d/M_{id*} \approx \sqrt{8/\pi}(\lambda_z/d)^3 \ll 1$.

Consider the effect of magnetic field direction variation on tunneling. Expression (21) determines tunneling matrix elements which mix different excitonic states. It follows from (21) that under increasing $\lambda_x^{-1}$ (i.e. under angle $\varphi$ between the field

**B** and growth axis $z$) the oscillation period of the phase multiplier in the integral grows. Hence the increase of $B_x$ suppresses tunneling. Let the magnetic length $\lambda = (\lambda_x^{-4} + \lambda_z^{-4})^{-1/4}$ be smaller than the exciton radius ($a_{ex}/\lambda \gg 1$). In the angle range satisfying the condition

$$\cos\varphi \gg \left(\frac{\lambda}{a_{ex}}\right)^2 \tag{30}$$

which is equivalent to (25) the wave functions can be expressed as follows:

$$\psi_d = \chi_0(\mathbf{r}, \mathbf{Q}) \quad ; \quad \psi_{id} = \chi_0\left(\mathbf{r}, \mathbf{Q} - \mathbf{e_y}\frac{d}{\lambda_x}\right)$$

$$\chi_0(\mathbf{r}, \mathbf{Q}) = e^{\frac{i}{2}\gamma \mathbf{Qr}} \phi_0\left(\mathbf{r} + \lambda_z^2 \mathbf{e}_z \times \mathbf{Q}\right)$$

After substitution of these functions in the integral (21) we obtain the expression for module of tunneling matrix element

$$|T_{e,h}| \approx t_{e,h} \exp\left(-\frac{d^2}{4\lambda^2}\frac{\sin^2\varphi}{\cos\varphi}\right) \tag{31}$$

Under the deviation of **B** direction from crystal growth axis the tunneling probability exponentially decreases and at the angles close to $\frac{\pi}{2}$ has the asymptotic

$$T_{e,h} = t_{e,h}\frac{4a_r}{a_d + a_{id}}\left[\left(\frac{m_{h,e}da_r}{M\lambda^2}\right)^2 \sin^2\varphi + 1\right]^{-3/2} \tag{32}$$

$$a_r = \frac{a_d a_{id}}{a_d + a_{id}}$$

Hence the minimal tunneling probability is realized at the orientation of the magnetic field **B** perpendicular to growth direction. If $m_{e,h}da_r/M\lambda^2 \gg 1$ is satisfied than $T_e, h$ decrease in inverse proportion to cube of field amplitude

$$T \propto \frac{1}{B^3}$$

Consider the solution of dispersion equation with tunneling being taken into account. It is convenient to introduce the following notations

$$\varepsilon_1(\mathbf{Q}) = E_d(\mathbf{Q})$$
$$\varepsilon_2(\mathbf{Q}), \Delta(\mathbf{Q}) = \frac{1}{2}\left[E_{id}\left(\mathbf{Q} + \mathbf{e}_y\frac{d}{\lambda_x^2}\right) \pm E_{id}\left(\mathbf{Q} - \mathbf{e}_y\frac{d}{\lambda_x^2}\right)\right] \tag{33}$$

Dispersion equation can be written in the following form

$$\left[(\varepsilon - \overline{\varepsilon})^2 - E_+^2\right]\left[(\varepsilon - \overline{\varepsilon})^2 - E_-^2\right] = (\Delta + edF)^2(\varepsilon - \varepsilon_1)^2 \tag{34}$$

here

$$\begin{aligned}E_\pm(\mathbf{Q}) &= \sqrt{\xi^2(\mathbf{Q}) + T_\pm^2} \\ \overline{\varepsilon}, \xi &= \frac{1}{2}\left[\varepsilon_1(\mathbf{Q}) \pm \varepsilon_2(\mathbf{Q})\right] \quad , \quad T_\pm = |T_e| \pm |T_h| \end{aligned} \quad (35)$$

Equation (34) determines exciton energies measured from the middle of the distance between electron and hole energy levels $\varepsilon_g$ in an isolated QW. This equation can be resolved exactly under the realization of the condition $\Delta(\mathbf{Q}) = -edF$. Corresponding solution let's denote as $\Omega_{1,2}^\pm$. Two lowest levels

$$\Omega_1^\pm = \overline{\varepsilon} - E_\pm \quad (36)$$

are generically coupled with two states of direct excitons splitted into symmetrical and asymmetric combinations. Two highest levels

$$\Omega_2^\pm = \overline{\varepsilon} + E_\pm \quad (37)$$

are composed mainly of indirect exciton states. The solution of eq. (34) has simple geometrical interpretation. The condition for the existence ($\Delta = -edF$) of the solution determines the intersection line in the plane $(Q_x, Q_y)$ of two indirect exciton spectrum branches (shifted by $\pm \mathbf{e}_y d/\lambda_x^2$). At zero electric field it takes place at the line $Q_y = 0$. These states (with $Q_y = 0$) are optically active and manifest themselves in optical experiments. At $F = 0$ the states $\Omega_{1,2}^-$ are antisymmetric in respect to $z$ and only transitions from the ground $\Omega_1^+$ and highest $\Omega_2^+$ levels are allowed, i.e. only lines corresponding to frequencies

$$\omega_{1,2} = \varepsilon_g + \overline{\varepsilon}(0) \pm \sqrt{\xi^2(0) + T_+^2}$$

will be observed. The behavior of the ground state $\Omega_1^+$ under the increase of the transverse component of the magnetic field is quite different from that of the excited state $\Omega_2^+$. Most representative in this respect is the case of $B_z = 0$ (the field is perpendicular to growth direction). In this case

$$\begin{aligned}\overline{\varepsilon}(\mathbf{Q}) &= -\frac{\varepsilon_d + \varepsilon_{id}}{2} + \frac{d^2}{4M\lambda^4} + \frac{Q^2}{2M} \\ \xi(\mathbf{Q}) &= \frac{\varepsilon_d - \varepsilon_{id}}{2} + \frac{d^2}{4M\lambda^4}\end{aligned} \quad (38)$$

Application of the electric field $F$ parallel to growth direction results in the asymmetry of excitonic spectrum in respect to momentum $\mathbf{Q}$. The extrema located initially (at $F = 0$) at $Q = 0$ shifts from the coordinate origin. It results in the shift of absorption spectra lines quadratic in respect to electric field. The ground state level lifts down

$$\Omega_1^+ = \Omega_1^+(0) - e^2 d^2 F^2 \frac{(E_+ - \xi)^2}{8 E_+ T_e T_h}$$

In the case of strong field (weak tunneling) this shift is very small. The shift of the excited level is much more pronounced

$$\Omega_2^+ = \Omega_2^+(0) + e^2 d^2 F^2 \frac{(E_+ + \xi)^2}{8 E_+ T_e T_h}$$

This effect is straightforward because direct exciton states are insensitive to electric field in our approximation (under the neglect of intrawell Stark-effect). In weak electric field $F$ the shifts of all 4 spectrum branches in the vicinity of $Q = 0$ can be found with the help of perturbation theory. At $B_z = 0$ ($\varphi = \pi/2$) from (38) and (34) it follows

$$\delta\Omega_1^\pm = \frac{m_1^\pm \mp M}{2 M m_1^\pm}\left(Q + \frac{M}{M \mp m_1^\pm} Q_0\right)^2 + \frac{Q_0^2}{2(M \mp m_1^\pm)}$$

$$\delta\Omega_2^\pm = \frac{m_2^\pm \pm M}{2 M m_2^\pm}\left(Q + \frac{M}{M \pm m_2^\pm} Q_0\right)^2 + \frac{Q_0^2}{2(M \pm m_2^\pm)}$$

where the notations are introduced

$$Q_0 = e M \lambda^2 F \quad ; \quad \frac{1}{m_{1,2}^\pm} = \frac{d^2}{4 M \lambda^4}\frac{(E_\pm \mp \xi)^2}{8 E_\pm T_e T_h}$$

The solutions of the dispersion equation (34) for the transverse direction of **B** are presented in Fig. 1. Fig. 1a corresponds to the absence of the electric field (F = 0) and symmetrical spectrum. Energy spectrum at the electric field $F = (\varepsilon_d - \varepsilon_{id})/ed$ is presented in Fig. 1b. The most characteristic feature of energy spectrum in the transverse magnetic field is the existence of two symmetrically located (at $Q = \pm d/\lambda_x^2$) extrema in the ground state branch. The appearance of these two minima results from tunneling mixing of direct and indirect (in **r**-space) excitonic states. The electric field doesn't effect practically the energy position of the central minimum while one of the side extrema lifts down proportionally to $F$. The energy of side extremum meats to the central one at the field $edF_c \approx \varepsilon_d - \varepsilon_{id}$. At further increase of $F$ the side minimum corresponding to indirect both in **k**- and **r**-space excitons becomes the ground state minimum. In experiment it will manifest itself as quenching of ground state luminescence at $F > F_c$. Photoluminescence intensity will decrease in proportion to central minimum population $I \propto \exp\{-ed(F - F_c)/kT\}$. This effect is much more pronounced than **r**-space indirect exciton photoluminescence intensity reduction in comparison with **r**-space direct exciton. At **k** = 0 photoluminescence intensity of an indirect exciton is reduced only as a result of wave-function overlap reduction. At **k** $\neq$ 0 exciton optical recombination is strictly forbidden because of momentum conservation. The regime of magnetic field induced photoluminescence quenching can be experimentaly observed if the energy separation $\delta E$ between ground state exciton minimum at **k** $\neq$ 0 and exciton energy at **k** = 0 is greater than the temperature. In the limit $\omega_H \equiv d^2/M\lambda^4 > \varepsilon_d - \varepsilon_{id}$ the minimum at $\mathbf{k} = 0$ corresponds to **r**-space direct excitons ($\delta E = ed(F - F_c)$) and the condition for the possibility of experimental observation of magnetic field induced photoluminescence quenching coincides with the condition for the observation of **r**-space indirect excitons in the absence of magnetic field. For example for DQW structure with the parameters $GaAs/Al_{0.3}Ga_{0.7}As/GaAs$ : $30\text{Å}/40\text{Å}/30\text{Å}$ at $B = 20T$ we have $\omega_H \approx 20meV$ that is greater than experimentaly observed values of exciton binding energy. Note that **r**-space indirect exciton photoluminescence was clearly resolved in DQW with $38.2\text{Å}$ barrier width [14].

# 5 Conclusion

We have shown that magnetoexciton spectrum of DQWS can be explicitly calculated in the tight-binding basis at arbitrary orientation of magnetic field in respect to growth direction. In asymmetric DQWS the spectrum is asymmetric in momentum. It corresponds to the general symmetry properties of the system which is characterized by both time-reversal and space-inversion symmetries. Hence a linear in momentum term in the energy becomes allowed. The physical origin of this term because of its symmetry nature can be understood on classical grounds. Electron and hole comprising moving exciton (momentum $\mathbf{k} \neq 0$) experience Lorentz force in magnetic field which shifts electron and hole in opposite directions and create a net dipole moment of exciton [12]

$$\mathbf{d} = \frac{c}{H^2}\mathbf{k} \times \mathbf{H}$$

The dipole moment $\mathbf{d}$ interacts with internal effective electrical field $\mathbf{E}_{eff}$ which is nonzero in asymmetric structure. So the linear in momentum term in the energy can be interpreted as the term describing interaction of exciton dipole moment induced by magnetic field $\mathbf{d}$ with internal effective electrical field $\Delta E = \mathbf{dE}_{eff}$.

The excitonic energy as a function of total momentum contains two pairs of minima differing in the value of exciton total momentum. At small asymmetry the lowest and the highest minima correspond to almost direct in k-space optically active excitons while the intermediate two correspond to essentially indirect excitons. The degree of spectrum asymmetry can be varied by external electrical and magnetic fields. At large enough asymmetry (in coupled QWs of second type) the exciton state with nonzero momentum can become a ground state. This state corresponds to the exciton indirect in **k**- as well as in **r**-space. In experiment the crossover from **k**-space direct to **k**-space indirect excitons will be accompanied by luminescence quenching. As is well known in bulk materials in crossed electrical and magnetic fields oscillator strength is exponentially reduced. It is shown in the paper that in ADQWS the two regimes exist corresponding to both (almost direct and indirect) types of excitonic ground state. Oscillator strength reduction and luminescence quenching become appreciable only at strong enough asymmetry in high fields.

# Acknowledgments

This work has been supported in part by Russian Basic Research Foundation, Russian Program "Physics of Solid State Nanostructures" and INTAS Grant 93-1704-ext. The work of one of us (I.V.T.) was supported in part by INTAS Grant 93-2492-ext within the research program of International Center for Fundamental Physics in Moscow.

# Figure captions

Figure 1. The qualitative dependence of exciton energy on $y$-component of momentum: $(a)$ - in the absence of electric field $F$; $(b)$ - in the presence of electric field $F = (\varepsilon_{id} - \varepsilon_{id})/ed$.